\documentclass[aps,prb,epsfigm,twocolumn,showpacs]{revtex4-1}
\usepackage{graphicx}
\usepackage{amsmath}
\usepackage{amsfonts}

\begin{document}

\title{Symmetry induced hole-spin mixing in quantum dot molecules}

\author{Josep Planelles, Fernando Rajadell, Juan I. Climente}
 \affiliation{Departament de Qu\'{\i}mica F\'{\i}sica i Anal\'{\i}tica, 
 Universitat Jaume I, E-12080, Castell\'o, Spain}
 \email{josep.planelles@uji.es}
 \homepage{http://quimicaquantica.uji.es/}
\date{\today}

%\maketitle

\begin{abstract}
We investigate theoretically the spin purity of single holes confined in vertically coupled 
GaAs/AlGaAs quantum dots (QDs) under longitudinal magnetic fields. 
A unique behavior is observed for triangular QDs, by which the spin is largely pure 
when the hole is in one of the dots, but it becomes strongly mixed when an electric field is
used to drive it into molecular resonance.
The spin admixture is due to the valence band spin-orbit interaction, which is greatly 
enhanced in $C_{3h}$ symmetry environments. The strong yet reversible electrical control of 
hole spin suggests that molecules with $C_3$-symmetry QDs, like those obtained with [111] growth,  
 can outperform the usual $C_2$-symmetry QDs obtained with [001] growth 
for the development of scalable qubit architectures.\\
\end{abstract}


 \maketitle

%%%%%%%% \section{Introduction} %%%%%%%%

Single spins confined in III-V semiconductor QDs are currently
considered as potential qubits for solid-state quantum information processing,
which combine fast optical and electrical manipulation with prospects of scalability.\cite{LossARCMP,WarburtonNM,DeGreveRPP,RamsaySST}
In the last years, heavy hole spin qubits have emerged as a robust and 
long-lived alternative to electron spins, as they can be less sensitive to dephasing
from nuclear spins.\cite{WarburtonNM,LossARCMP,DeGreveRPP,RamsaySST,DahbashiPRL,HouelPRL,CoishPRL,FrasPRB,HeissPRB}
Significant advances have been reported on hole spin initialization, control and 
readout by means of optical excitations\cite{WarburtonNM,DeGreveRPP,RamsaySST,MuellerPRB,GoddenPRL,GoddenPRB,GreilichNP}, % ,RamsayPRL},
and different proposals for electrical control have been put forward.\cite{SzumniakPRL,BudichPRB,PingenotPRB,JovanovPRB,DotyPhys}

Although most works so far have focused on QDs grown along [001], it has been 
noted that the $C_{2}$ point symmetry of such systems gives rise to a splitting of bright exciton states
which limits the fidelity of optical hole spin preparation.\cite{GoddenPRB,GoddenAPL}
No such splitting is however expected in [111] grown QDs owing to their higher ($C_{3}$) symmetry,\cite{KarlssonPRB,SinghPRL}
which hence become an alternative worth exploring.
Early studies on single (In)GaAs/AlGaAs QDs grown along [111] have revealed that hole states 
have weak heavy hole-light hole (HH-LH) coupling due to the large aspect ratio, which is a prerequesite to obtain pure hole spins
and minimize the impact of hyperfine interaction with the lattice nuclei.\cite{KarlssonAPL}
In turn, magneto-photoluminescence spectra have reported characteristic differences from [001] grown QDs
which were ascribed to the influence of the $C_{3}$ symmetry on the hole states.\cite{SallenPRL}

%their magneto-photoluminescence spectra revealing characteristic differences from their [001] counterparts,
%which may be ascribed to a magnetic field induced hole spin mixing mechanism absent in lower symmetry QDs.

In this paper, we move forward and study hole states confined in quantum dot molecules (QDM) formed by
a pair of vertically stacked QDs grown along [111]. QDMs present several advantadges over single QDs for 
qubit development, including readout independency from initialization and measurement protocols\cite{VamivakasNAT},
higher fidelity of spin preparation\cite{MuellerPRB} and enhanced wavelength tunability with external electric fields, 
which greatly improves prospects of scalability.\cite{EconomouPRB}
We consider [111] grown GaAs/AlGaAs QDMs with triangular shape, similar to those reported in Refs.~\cite{ZhuSMALL,MicheliniX}, 
adding longitudinal magnetic and electric fields to control the Zeeman splitting and charge localization.
Our calculations show that the HH spin purity is high when the hole is confined in individual QDs,
but severe spin admixture takes place when the electric field is used to form molecular orbitals.  
The spin admixture follows from the formation of orbitals with approximate $C_{3h}$ point group symmetry, 
%resulting from the triangular lateral confinement and the resonant electric field, 
 which enables otherwise forbidden spin-orbit interactions (SOI). 
The symmetry-induced SOI does not mix nearby Zeeman sublevels, but it couples bonding
and antibonding molecular states split by the tunneling energy. 
This is in sharp contrast with usual [001] grown QDMs, with $C_{2h}$ symmetry, 
where tunneling is normally a spin-preserving process.\cite{StinaffSCI}

Since the activation of SOI mechanisms is generally associated with a descent of the system symmetry
(e.g. system and bulk inversion asymmetry\cite{Winkler_book}, QDM misalignment\cite{DotyPRB}),
the enhancement of SOI for $C_3$ QDs is apparently counterintuitive.
Yet, we observe strong spin admixture between hole states over 1 meV apart, which is $2.5$ times 
greater than the largest spin-orbit anticrossing measured in [001] grown QDMs.\cite{DotyPRB} 
We provide an explanation through group theory analysis of the the multi-band k$\cdot$p Hamiltonian for holes,
showing this is an exclusive property of $C_3$ systems, and discuss the implications of these findings for 
the development of hole spin qubit architectures.

%%%%%%%% \section{Theory} %%%%%%%%

The Hamiltonian we use to describe hole states reads:
\begin{equation}
{\mathbb H} =  {\mathbb H}_{BF} + {\mathbb H}_{B} + {\mathbb H}_{strain} + \left( V(\mathbf{r})\, + \,e\,(\phi_{pz}(\mathbf{r}) - F\,z)\right) \,\mathbb{I}.
\label{eq:H}
\end{equation}
Here ${\mathbb H}_{BF}$ is the four-band Burt-Foreman Hamiltonian\cite{foreman} for [111]
grown zinc-blende crystals, which considers HH-LH subband coupling as in the Luttinger model\cite{luttinger}
but including position-dependent effective masses. %\cite{Voon_book}
${\mathbb H}_{B}$ represents the terms coming from a magnetic field applied along the growth ($z$) direction.
${\mathbb H}_{strain}$ is the strain Hamiltonian,
$V(\mathbf{r})$ the band-offset potential, $e$ the hole charge, $\phi_{pz}$
the piezoelectric potential, $F$ an axial electric field and $\mathbb{I}$ a rank-4 identity matrix. 
Further details on the Hamiltonian can be found in the Supplemental Material.\cite{supmat}
Hamiltonian (\ref{eq:H}) is solved numerically after obtaining the strain
tensors and piezoelectric fields using the \emph{Comsol} package.
The eigenstates are Luttinger spinors of the form:
\begin{equation}
|n\rangle =\,\sum_{Jz=-3/2}^{3/2}\,f_{J_z}^n(\mathbf{r})\,|J=3/2,J_z\rangle,
\label{eq:spinor}
\end{equation}
\noindent where $f_{J_z}^n(\mathbf{r})$ is the envelope function associated to
$|J=3/2,J_z\rangle$, the periodic function with Bloch angular momentum
$J_z$. $J_z=\pm 3/2$ correspond to spin up and spin down HH components,
while $J_z=\pm 1/2$ correspond to LH components. The expectation value
$\langle J_z \rangle = \sum_{J_z} J_z \langle f_{J_z}^n | f_{J_z}^n \rangle$ 
can be taken as a measure of the hole spin purity, with
$\langle J_z \rangle \approx \pm 3/2$ indicating nearly pure HH spin up ($\Uparrow$)
or spin down ($\Downarrow$) states.

\begin{figure}[h]
\includegraphics[width=0.5\textwidth]{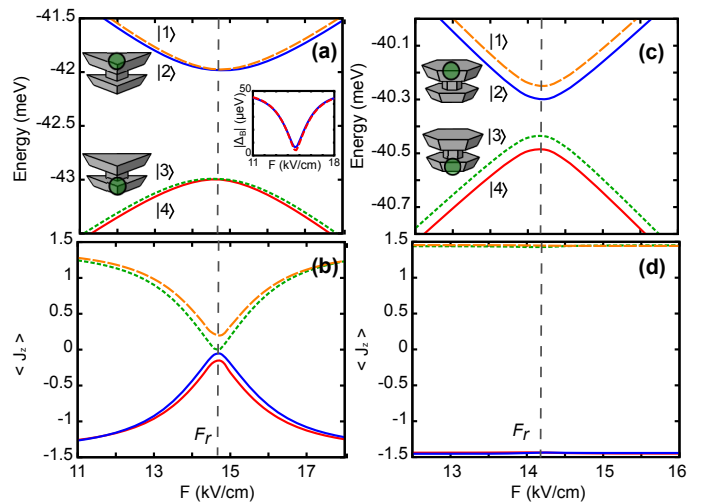}
\caption{(a) and (b): hole energy levels and Bloch angular momentum expectation value of a triangular QDM
grown along [111], as a function of a vertical electric field.  
(c) and (d): same but for a hexagonal QDM. 
Note the strong spin mixing for triangular QDMs near the resonant field $F_r$.
%Energies are referred to the top of the valence band. 
 The structures in (a) and (c) show the hole localization for each doublet.
The inset in (a) shows the Zeeman splitting of the upper 
(solid line) and lower (dashed line) doublet.}\label{fig1}
\end{figure}

For our calculations we consider pyramidal GaAs/Al$_{0.3}$Ga$_{0.7}$As QDMs with triangular confinement, 
similar to those obtained by metallorganic vapor deposition.\cite{ZhuSMALL,MicheliniX}
The vertically stacked QDs are separated by a barrier of thickness $d=2$ nm, although they are interconnected by a 
thin Al$_{0.05}$Ga$_{0.95}$As vertical quantum wire which enhances tunnel coupling -see structure in Fig.~\ref{fig1}(a)-.\cite{supmat} 
A weak magnetic field of $B=0.2$ T is applied along the coupling direction.
Figure \ref{fig1}(a) shows the energy of the first four hole states under the influence of
a vertical electric field $F$. At $F=11$ kV/cm we see two Zeeman-split doublets. 
The upper one (states $|1\rangle$ and $|2\rangle$) corresponds to the main component of the hole spinor in the top QD, 
which is slightly bigger than the bottom QD. 
In turn, the lower doublet (states $|3\rangle$ and $|4\rangle$) has the main component in the bottom dot.
 Since the increasing electric field favors the occupation of the bottom dot,
a charge transfer anticrossing takes place at $F=14.7$ kV/cm (resonant electric field, $F_r$), 
where bonding and antibonding molecular orbitals are formed.
The behavior closely resembles that of [001] grown QDMs,\cite{StinaffSCI,ClimentePRB,DotyPRL}
except for one anomaly: the Zeeman splitting of both doublets is quenched near the resonant electric field,
see Fig.~\ref{fig1}(a) inset.

The Zeeman splitting suppression can be seen as a vanishing effective g-factor. %between spin $1/2$ states.
Unlike in previous reports, however, the origin of this effect cannot be ascribed 
to the different g-factor of the QD and barrier materials,\cite{DotyPRL2}
as the QDs and the vertical wire connecting them have similar composition. 
Because electrically tunable g-factors are of interest for spin manipulation,\cite{DotyPhys}
we further investigate into the origin of this phenomenon.
Fig.~\ref{fig1}(b) shows the hole Bloch angular momentum expectation value of the four states under consideration.
As can be seen, away from the resonant field, $\langle J_z \rangle$ gradually approaches $\pm 3/2$, 
indicating that the hole states confined in individual QDs are nearly HH states with fairly pure spin.
In the vicinity of $F_r$, however, $\langle J_z \rangle \approx 0$, which means that for molecular
states the spin becomes completely mixed.

For comparison, since [111] grown QDs normally have either triangular or hexagonal shape,\cite{ZhuSMALL,AbbarchiJCG,PfundAPL} 
in Fig.~\ref{fig1}(c-d) we study a QDM formed by hexagonal QDs. One can see that no spin mixing takes place near $F_r$ in this case.
Actually, the behavior is now the same as observed in vertically aligned [001] grown QDMs 
with $C_{2h}$ symmetry QDs, where tunneling is a spin preserving process.\cite{StinaffSCI} 
It follows that the strong spin mixing in Fig.~\ref{fig1}(b) is not a consequence of the [111] 
crystal orientation, but rather of the triangular envelope confinement. In fact, we also
observe it for triangles grown along [001], see Supplementary Material.\cite{supmat}
It does not result from the presence or absence of strain either,
as similar results are obtained using lattice mismatched materials such as InAs/GaAs.\cite{supmat}
Likewise, it is not induced by the magnetic field, as ${\mathbb H}_{B}$ 
is a diagonal term which does not couple different spinor component.\cite{supmat} % and the applied field $B$ is longitudinal.
It does not take place in single QDs either. 
It is an exclusive property of QDMs with triangular confinement. 

\begin{figure}[h]
\includegraphics[width=0.45\textwidth]{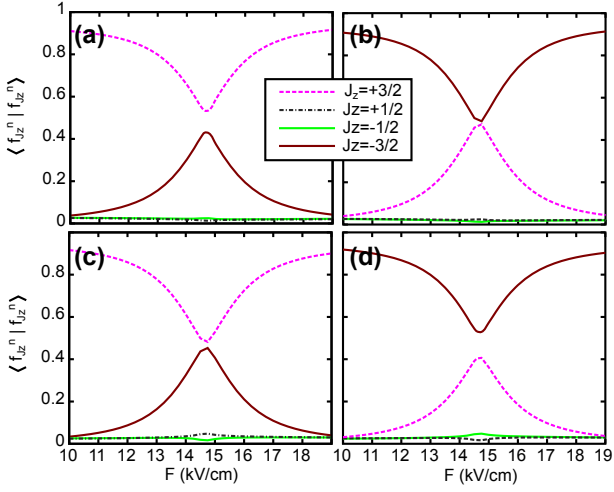}
\caption{Weight of the different Luttinger spinor components of the hole states in Fig.~\ref{fig1}(a), 
as a function of the electric field. Panels (a) to (d) correspond to states $|1\rangle$ to $|4\rangle$, 
respectively. The states are almost exclusively HH ($J_z=\pm 3/2$).}\label{fig2}
\end{figure}

Further insight into the hole spin mixing mechanism is obtained in Fig.~\ref{fig2}, which plots 
the weight of the four spinor components corresponding to each of the states $|1\rangle$ to $|4\rangle$ 
of the triangular QDM. 
Two conclusions can be extracted:
(i) LHs play a minor role in all cases, the mixing is essentially between 
HH components with orthogonal spin projections ($J_z=\pm 3/2$); 
(ii) states $|1\rangle$ and $|4\rangle$ (panels (a) and (d)) seem to exhibit
complementary behavior, and so do $|2\rangle$ and $|3\rangle$ (panels (b) and (d)).
This suggests that the spin mixing is due to independent interactions between 
$|1\rangle$ and $|4\rangle$ on the one hand, and $|2\rangle$ and $|3\rangle$ 
on the other.

To understand the origin of the HH spin mixing we resort to a point group 
theory analysis.  Eq.~(\ref{eq:H}) Hamiltonian can be simplified as 
$\mathbb{H} \approx {\mathbb H}_{LK}^{[111]} + {\mathbb H}_{B} 
+ \left( V(\mathbf{r})\, + e\,F\,z) \right) \,\mathbb{I}$, where
we have disregarded strain terms --which are weak for GaAs/AlAs heterostructures--
 and approximated ${\mathbb H}_{BF}$ by the (constant mass) Luttinger-Kohn Hamiltonian:
%In fact, in Equation (\ref{eq:H}), ${\mathbb H}_{BF}$ can be approximated by the
%(constant mass) Luttinger-Kohn Hamiltonian: %\cite{Voon_book} 
%
\begin{equation}
\label{eqb1}
 {\mathbb H}_{LK}^{[111]}=-\frac{1}{2} \left( 
\begin{array}{llll}
\hat{P} + \hat{Q} & -\hat{S} & \hat{R} & 0 \\
-\hat{S}^{\dag}& \hat{P} - \hat{Q} & 0 & \hat{R} \\
\hat{R}^{\dag}& 0& \hat{P} - \hat{Q} & \hat{S}  \\
0& \hat{R}^{\dag}& \hat{S}^{\dag}& \hat{P} + \hat{Q} 
\end{array} 
\right)
\end{equation}
\noindent with
\begin{equation}
\label{eqb2}
\begin{array}{lll}
\hat{P} \pm \hat{Q} &=& (\gamma_1 \pm \gamma_3) (k_x^2 + k_y^2) + (\gamma_1 \mp 2 \gamma_3) k_z^2 \cr
\hat{R} &=& -\frac{1}{\sqrt{3}} (\gamma_2+2 \gamma_3) k_-^2  + \frac{2\sqrt{2}}{\sqrt{3}} (\gamma_2-\gamma_3) k_+ k_z \cr
\hat{S} &=& -\frac{\sqrt{2}}{\sqrt{3}} (\gamma_2-\gamma_3)  k_+^2 + \frac{2}{\sqrt{3}} \,(2 \gamma_2+\gamma_3) k_- k_z \cr
\end{array}
\end{equation}
\noindent where $\gamma_1,\gamma_2,\gamma_3$ are the Luttinger parameters and $k_{\pm}=k_x\pm ik_y$.
%For many III-V materials, like GaAs, $\gamma_2 \approx \gamma_3$ and one can 
%replace them both by $ \bar \gamma=(\gamma_2+\gamma_3)/2$ 
%in the $\hat{R}$ and  $\hat{R}^{\dag}$ matrix elements.
%Then, $\mathbb{H}_{LK}$ has axial symmetry.\cite{SercelPRB}  
%The same occurs for the strain terms. 
 One can then see that $\mathbb{H}$ has $C_3$ point symmetry set by the confining potential $V(\mathbf{r})$.
%$ {\mathbb H}_{LK}^{[111]}$    % has rotational symmetry, but the symmetry of $\mathbb{H}$ is lowered to $C_{3v}$ by the 
%including confining potential  $V(\mathbf{r})$ and the axial magnetic field, has  $C_3$ rotational symmetry. 
%
Near the resonant electric field, however, an additional approximate symmetry must be considered.
Even if the two QDs forming the QDM are not identical, the electric field restores an
\emph{effective} parity symmetry, the bonding and antibonding HH molecular orbitals forming even and
odd functions with respect to a mirror plane in between the QDs.\cite{ClimentePRB}
The corresponding point group is then $C_{3h}$. Note that for $ {\mathbb H}_{LK}^{[111]}$ to
hold exact $C_{3h}$ symmetry, we need to impose the so-called axial approximation ($\gamma_2=\gamma_3$), 
which is actually valid for many III-V materials, such as GaAs. 
Actually, we do not impose exact symmetry in the numerical calculations and nevertheless,
the obtained results reveal, as expected, a high degree of symmetry.
\begin{figure}[h]
\includegraphics[width=0.45\textwidth]{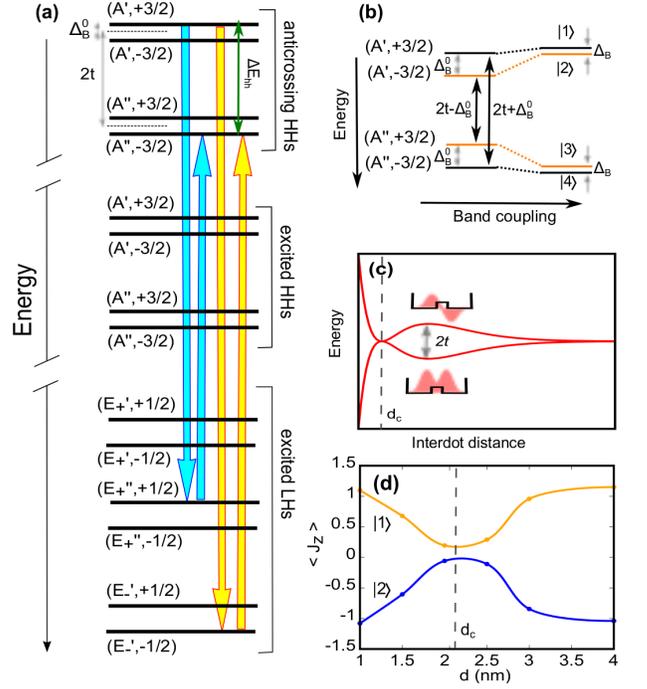}
\caption{(a) Diagram of single band hole energy levels under a longitudinal magnetic field for
a QDM with symmetry $C_{3h}$. The labels $(X^p,J_z)$ indicate the symmetry of the levels. 
$X$ represents the rotational symmetry of the envelope function ($A$, $E_\pm$) while 
$p=\prime$ or $\prime \prime$ represents the even/odd parity and $J_z$ indicates the non-zero component
of the spinor. Thick arrows denote the symmetry allowed level couplings for the ground state.
(b) Energy structure of anticrossing hole states before and after switching on off-diagonal terms in $\mathbb{H}_{LK}^{[111]}$.
(c) Typical dissociation energy spectrum of holes in QDMs at $B=0$ T, showing
a bonding-antibonding ground state reversal at $d_c$.
(d) Calculated spin purity of states $|1\rangle$ and $|2\rangle$ for different
interdot distances at the resonant electric field. Spin mixing is strongest around
$d_c$.}\label{fig3} 
\end{figure}

The anticrossing of Fig.~\ref{fig1}(a) can be rationalized considering the symmetry of
the hole spinors in the double group $\bar{C}_{3h}$. %In the vicinity of $F_r$, 
%when the energy levels are closer and mixing should be maximized, the double group is $\bar{C}_{3h}$. 
Within this group $|1\rangle$ and $|4\rangle$ have $E_{-3/2}$ symmetry, while 
$|2\rangle$ and $|3\rangle$ have $E_{3/2}$ symmetry.\cite{supmat} 
 The different symmetry of $|1\rangle$ and $|2\rangle$ ($|3\rangle$ and $|4\rangle$)
explains the lack of interaction within the Zeeman doublets,
inspite of the quasi-degeneracy. It also becomes clear why $|1\rangle$
and $|4\rangle$ ($|2\rangle$ and $|3\rangle$) interact separately, as observed in
Fig.~\ref{fig2}.

The above picture differs from the widely studied [001] grown QDMs with circular confinement, 
where the symmetry of all four spinorial states involved in the molecular anticrossing is 
different\cite{ClimentePRB}, which results in the absence of interaction and hence 
spin-preserving tunneling. %\cite{StinaffSCI} 
Sizable hole spin mixing has been observed only in QDMs with significant misalignment\cite{DotyPRB}, 
because the symmetry is then completely reduced ($\bar{C}_1$ point group). 
Nevertheless, the largest spin anticrossing measured for such system is $0.4$ meV,
corresponding to InAs QDMs with anomalously large lateral offset.\cite{EconomouPRB} 
 This is $2.5$ times smaller than the $1$ meV gap we estimate in Fig.~\ref{fig1},
and $6$ times smaller than the $2.5$ meV we predict for InAs/GaAs QDMs.\cite{supmat}
 
To explain the unusual strength of the spin-orbit coupling in triangular QDMs,
we can examine the envelope symmetry of ${\mathbb H}_{LK}^{[111]}$
(within axial approximation) and the ensuing eigenfuctions.
In the $C_{3h}$ point group, the terms of Eq.~(\ref{eqb1}) form basis of the 
following irreducible representations:
\begin{equation}
\label{eqc5}
\Gamma_{H_{LK}}=
\left( 
\begin{array}{llll}
A^{\prime} & E_-^{\prime \prime} & E_+^{\prime} & 0 \\ 
E_+^{\prime \prime} & A^{\prime} & 0 & E_+^{\prime} \\ 
E_-^{\prime} & 0 & A^{\prime} & E_-^{\prime \prime} \\ 
0 & E_-^{\prime } & E_+^{\prime \prime} & A^{\prime} 
\end{array} \right).
\end{equation}
A remarkable consequence is that for any hole state $|n\rangle$, the envelope functions 
of the spin up and down HH components, $f_{3/2}^n$ and $f_{-3/2}^n$ in Eq.~(\ref{eq:spinor}), 
must have the same symmetry except for the even/odd parity (e.g. $A^\prime$ and $A^{\prime \prime}$).
 This is a key factor in determining the strength of the spin admixture, as we show in the 
perturbative analysis below.

Hamiltonian $\mathbb{H}$ can be split as a sum of diagonal and off-diagonal terms,
$\mathbb{H}=\mathbb{H}_0 + \mathbb{H}^\prime$, the latter being responsible for band coupling.
If we disregard $\mathbb{H}^\prime$, the levels anticrossing at $F_r$ are those represented at the top 
of Fig.~\ref{fig3}(a), namely two Zeeman doublets formed by HHs with opposite spin 
($J_z=\pm 3/2$) but the same rotational symmetry ($A$) and parity ($^\prime$ or $^{\prime \prime}$). 
Each doublet is split by a Zeeman term $\Delta_B^0$ and separated from each other by an amount $2t$, 
where $t$ is the HH tunneling integral. 
 Considering $\mathbb{H}^\prime$ as a perturbative term, the mixing between the spin up ground state
$|k^{(0)}\rangle=(A^\prime,+3/2)$ and any spin down HH state $|i^{(0)}\rangle$ is given by:
\begin{equation}
\label{pert}
|k^{(2)}\rangle =\sum_{i\neq k} \left(\sum_{j\neq k} \frac{\langle i^{(0)}|\mathbb{H}^\prime|j^{(0)}\rangle}{E_k^0-E_i^0}
\frac{\langle j^{(0)}|\mathbb{H}^\prime|k^{(0)}\rangle}{E_k^0-E_j^0} \right) |i^{(0)}\rangle 
\end{equation}
\noindent where $|j^{(0)}\rangle$ is the $j$-th intermediate state and $E^0_j$ its corresponding energy.
Notice that the strength of coupling is inversely proportional to $\Delta E_{hh} = E_k^0-E_i^0 $, i.e. the energy 
difference between the spin up and down HH states.
 As explained with detail in the Supplemental Material,\cite{supmat} 
 the symmetry of $\mathbb{H}^\prime$ operators, off-diagonal terms of Eq.(\ref{eqc5}), translates into selection rules which 
make the numerator of Eq.~(\ref{pert}) vanish for all except the two paths plotted with thick arrows in Fig.~\ref{fig3}(a).
Both paths involve excited LHs as intermediate states\cite{gamma2}, and the spin down HH is $|i\rangle=(A^{\prime\prime},-3/2)$, i.e.
a state participating in the molecular anticrossing. This implies $\Delta E_{hh}$ is small (few meV at most), 
and is in contrast with other point symmetries, where selection rules lead to coupling with higher excited HH states,
so that $\Delta E_{hh}$ is much larger.
%
%the switch on of the band coupling, i.e., the accounting for the extradiagonal $R$ and $S$ Hamiltonian matrix elements,
%enables the coupling of HH $\Uparrow$ with HH $\Downarrow$ mediated by LHs 
%(e.g. $\langle HH\!\Uparrow| S | LH\!\Uparrow\rangle \langle LH\!\Uparrow| R | HH\! \Downarrow \rangle$). Since $R$ and $S$
%have different parity, $|HH \! \Uparrow\rangle $ and $|HH \! \Downarrow\rangle$ must have different parity too. 
%Therefore, as indicated in Fig.~\ref{fig3}(b), the coupling occurs between states belonging to different doublets.
%In addition, because the energy splitting between the innermost states ($2t-\Delta_B^0$)
%is smaller than that between the outermost ones ($2t+\Delta_B^0$), the 
%interaction is stronger. This leads to an effectively reduced Zeeman splitting, $\Delta_B$,
%in agreement with the numerical results of Fig.~\ref{fig1}(a).
%
%It is worth noting that the strength of the spin mixing we observe is made possible
%by the $C_3$ symmetry. 
%
For example, if we consider QDMs with circular QDs (point group $C_{\infty h}$), the HH 
components coupled by ${\mathbb H}^{\prime}$ no longer have the same rotational symmetry, 
but they differ by three quanta of azimuthal angular momentum $M_z$.\cite{ClimentePRB}
Consequently, there is no coupling between the $M_z=0$ HHs forming the molecular 
anticrossing.\cite{supmat}
Having a QDM structure is also essential, as then $A^\prime$ and $A^{\prime\prime}$ are 
roughly split by the tunneling energy $2t$, which can be made small enough for the 
SOI to be efficient. By contrast, in single QDs the strong vertical confinement would
lead to several meV splitting. %A detailed perturbation analysis is given in Ref.~\cite{supmat}.

% tmp *** falta Zeeman suppression
The suppression of the Zeeman splitting observed in Fig.~\ref{fig1}(a) can be also understood
from the perturbative analysis. As indicated in Fig.~\ref{fig3}(b), the band coupling occurs between 
HH states belonging to different doublets. 
%Each doublet is split by a Zeeman term $\Delta_B^0$ and separated from each other by an amount $2t$, 
%where $t$ is the HH tunneling integral. 
 Because $\Delta E_{hh}$ is smaller for the innermost states ($2t-\Delta_B^0$) than for the outermost ones 
($2t+\Delta_B^0$), the interaction is stronger, leading to an effectively reduced Zeeman splitting, $\Delta_B$.

It is clear from the discussions above that tunneling must be an important parameter 
to control the strength of the spin mixing. One might then expect that spin mixing 
is enhanced for long interdot distances $d$, when tunneling energy $t$ is small.
Fig.~\ref{fig3}(d) shows the spin purity of the ground state HH for QDMs with different $d$
at resonant electric field.  Interestingly, the maximum spin mixing is found at 
intermediate distances, $d \approx 2$ nm.
This follows from the characteristic, non-monotonous decay of hole tunneling in 
QDMs.\cite{ClimentePRB,DotyPRL,JaskolskiPRB,BesterPRL}
As shown in the schematic of Fig.~\ref{fig3}(c), there is a critical distance $d_c$ 
where bonding and antibonding hole states are reversed.
At this point, $t$ has a relative minimum combined with large wave function delocalization,
which enables the strong spin mixing.
For $d < d_c$, $t$ increases rapidly, reducing the interaction. % and then, the quenching of the Zeeman splitting (see Fig. 3c and 3d). %% nota nacho: no mostrem zeemans en Fig. 3d
For $d > d_c$, $t$ eventually decreases but 
%% nota nacho: crec que la interpretacio que havia donat ades no es bona: estem sempre en Fr, ha d'haver wf en els 2 QD
%the envelope functions localize in opposite QDs 
%since the artificial resonance between QDs produced by the electric field is lost, 
%so that the possible interacting states display a reduced admixture
%and the Zeeman splitting is also recovered.
 so does the wave function delocalization. 
As a result we gradually retrieve the single QD limit, were spin mixing is weak.

%For $d < d_c$, $t$ increases rapidly and so does $\Delta_{hh}$ in Eq.~(\ref{eq:pt}).
%For $d > d_c$, $t$ eventually decreases but so do the matrix elements in Eq.~(\ref{eq:pt}),
%as the coupled $f_{-3/2}^n$ and $f_{3/2}^n$ envelope functions tend to localize in opposite QDs.

Electrical control of hole spins in QDMs has been proposed as a key ingredient 
for scalable qubit architectures.\cite{EconomouPRB}
So far, however, only [001] grown QDMs have been considered, where the main source of 
spin mixing was misaligment between the vertically stacked dots,\cite{DotyPRB}
which is a difficult parameter to regulate experimentally.
The $C_{3h}$-symmetry-induced spin mixing described here arises as a more robust and manageable 
mechanism.
It can also help increase the fidelity of spin control gate operations,
as this requires the spin mixed states to (i) be energetically well resolved, 
and (ii) be able to form indirect excitons with large optical dipole strength.\cite{EconomouPRB}
As for (i), we predict strong mixing between states 1 meV away from each other, 
larger than any previous measurement. As for (ii), unlike in misaligned QDMs, the spin 
mixing we describe is strong at the resonant electric field, 
 where direct and indirect excitons have comparable optical strength.
Another advantadge is the possibility to use weak magnetic fields, which limits the 
influence of the g-factor inhomogeneity of different QDs in the qubit scaling.

In summary, we have shown that triangular QDs can be used to build QDMs with
electrically controllable hole spin. The hole spin is well defined inside the individual
QDs, but the formation of delocalized  molecular orbitals with $C_{3h}$ symmetry enables SOI induced mixing
with unprecedented strength. The reversible control, the strength of the interaction
and the robust nature of the spin mixing mechanism imply that holes in triangular QDMs, 
like those obtained with [111] growth, form a promising system for quantum information 
processing with some advantadges as compared to circular [001] grown QDMs.

Support from UJI project P1-1B2014-24 and MINECO project CTQ2014-60178-P is acknowledged.

%\bibliography{holetrianglebib}{}

\begin{thebibliography}{10}
\bibitem{gamma2}
$\mathbb{H}^\prime$ has no term mixing spin up and down HH directly. The mixing is mediated by LH
through $\hat{R}$ and $\hat{S}$ operators. Thus, the spin admixture we report is proportional to the Luttinger parameters
$\gamma_2$ and $\gamma_3$.  Since the value of these parameters is modulated 
by the strength of the spin-orbit interaction, the extent of the spin admixture is ultimately due to spin-orbit interaction.
\end{thebibliography}
%\bibliographystyle{unsrt}

%\begin{thebibliography}{30}
%\end{thebibliography}

\end{document}

\title{Supplemental Material for: Symmetry induced hole-spin mixing in quantum dot molecules}

\author{Josep Planelles, Fernando Rajadell, Juan I. Climente}
 \affiliation{Departament de Qu\'{\i}mica F\'{\i}sica i Anal\'{\i}tica,
  Universitat Jaume I, E-12080, Castell\'o, Spain}
\email{josep.planelles@uji.es}
\homepage{http://quimicaquantica.uji.es/}
\date{\today}

\pacs{73.21.La,73.40.Gk,71.70.Ej}

\maketitle

%%%%%%%% \section{Theory} %%%%%%%%
\section{Theoretical model and parameters}

The Hamiltonian we use to describe hole states in QDs grown along the [001] direction is (atomic units):
\begin{equation}
{\mathbb H}^{[001]} =  {\mathbb H}_{BF}^{[001]} + {\mathbb H}_{B}^{[001]}  + {\mathbb H}_{strain}^{[001]} + \left( V(\mathbf{r})\, + \,e\,(\phi_{pz}^{[001]}(\mathbf{r}) - F\,z)\right) \,\mathbb{I}.
\label{eq:H}
\end{equation}
Here ${\mathbb H}_{BF}^{[001]}$ is the four-band Burt-Foreman Hamiltonian \cite{foreman}
for zinc-blende crystals, which considers HH-LH subband coupling
and position-dependent effective masses:
\small
\begin{eqnarray}
\label{equat1}
 H_{BF}^{[001]} & =& (\frac{1}{2} \sum_i^{x,y,z} k_i \frac{3L+M}{2} k_i) \mathbb I_0 - \sum_i^{x,y,z} k_i \frac{L-M}{3} k_i \, \mathbb J_i^2 + \frac{i}{3} (k_x (N-N') k_y- k_y (N-N') k_x) \, \mathbb J_z \nonumber \\
	     &  &  -\frac{1}{3} \sum_{i<j}^{x,y,z} (k_i (N+N') k_j + k_j (N+N')  k_i) \{\mathbb J_i,\mathbb J_j\} \nonumber \\
		  &  & +\frac{1}{6} [(k_+ (N-N') k_z-k_z (N-N') k_+) \,  \mathbb J_- + (k_z (N-N') k_- - k_- (N-N') k_z) \,  \mathbb J_+]
\end{eqnarray}
\normalsize
\noindent with ${\mathbb  I}$ a rank-4 identity matrix, $\{A,B\}=\frac{1}{2}(AB+BA)$, $X_\pm =X_x\pm X_y$ and ${\mathbb  J}_i$ the angular momentum $i$-component matrix:
\begin{equation}
\label{equat2}
\mathbb J_x = \left( \begin{matrix} 0& \frac{\sqrt{3}}{2} & 0 & 0 \cr \frac{\sqrt{3}}{2} & 0 & 1 & 0 \cr 0 & 1 & 0 & \frac{\sqrt{3}}{2} \cr 0 & 0 & \frac{\sqrt{3}}{2}& 0\end{matrix}\right) \;\;\;\;\;\;\;\;
\mathbb J_y =\left( \begin{matrix}0 & -i \frac{\sqrt{3}}{2} & 0 & 0 \cr i \frac{\sqrt{3}}{2} & 0 & -i & 0 \cr 0 & i & 0 & -i  \frac{\sqrt{3}}{2} \cr 0 & 0 & i \frac{\sqrt{3}}{2} & 0\end{matrix}\right) \;\;\;\;\;\;\;\;
\mathbb J_z = \left( \begin{matrix}\frac{3}{2} & 0 & 0 & 0 \cr 0 & \frac{1}{2} & 0 & 0  \cr  0 & 0 & -\frac{1}{2} &0\cr 0 &  0 & 0 & -\frac{3}{2}\end{matrix}\right) 
\end{equation}

\noindent By inserting eq. \ref{equat2} in eq. \ref{equat1} we end up with the following matrix representation:

\begin{equation}
\label{eq1}
 {\mathbb  H}_{BF}^{[001]} = - \left( 
\begin{array}{llll}
\hat{P'} & \hat{S_-} & -\hat{R} & 0 \\ 
\hat{S_-}^{\dag} & \hat{P"} & -\hat{C} & -\hat{R} \\ 
-\hat{R}^{\dag} & -\hat{C}^{\dag}&  \hat{P"}^* & -\hat{S_+}^{\dag} \\ 
0 & -\hat{R}^{\dag} & -\hat{S_+} & \hat{P'}^* 
\end{array}
\right)
\end{equation}

\noindent where,
\begin{equation}
\label{eq2}
\begin{array}{lll}
P' &=& \frac{1}{2} (k_x (L + M) k_x + k_y (L+M) k_y + k_z (2 M) k_z) +  \frac{i}{2} (k_x (N-N') k_y - k_y  (N-N') k_x)
\cr
P"  &=& \frac{1}{6} (k_x (L + 5 M) k_x + k_y (L + 5 M) k_y + 2 k_z (2 L + M) k_z) + \frac{i}{6} (k_x (N-N') k_y - k_y  (N-N') k_x)
\cr
P"^*  &=& \frac{1}{6} (k_x (L + 5 M) k_x + k_y (L + 5 M) k_y + 2 k_z (2 L + M) k_z) - \frac{i}{6} (k_x (N-N') k_y - k_y  (N-N') k_x)
\cr
P'^*  &=& \frac{1}{2} (k_x (L + M) k_x + k_y (L+M) k_y + k_z (2 M) k_z) -  \frac{i}{2} (k_x (N-N') k_y - k_y  (N-N') k_x)
\cr
R  &=& \frac{1}{2 \sqrt{3}} [k_x (L - M) k_x - k_y (L - M) k_y - i (k_x (N+N') k_y + k_y (N+N') k_x)]
\cr
R^{\dag} &=& \frac{1}{2 \sqrt{3}} [k_x (L - M) k_x - k_y (L - M) k_y + i (k_y (N+N') k_x+k_x (N+N') k_y)]
\cr
S_-  &=& - \frac{1}{\sqrt{3}}  [k_- N k_z + k_z N' k_-]
\cr
S_-^{\dag}  &=& - \frac{1}{\sqrt{3}}  [k_z N k_+ + k_+ N' k_z]
\cr
S_+  &=& - \frac{1}{\sqrt{3}} [k_+ N k_z + k_z N' k_+]
\cr
S_+^{\dag}  &=& - \frac{1}{\sqrt{3}} [k_z N k_- + k_- N' k_z]
\cr
C  &=& -\frac{1}{3} (k_z (N-N') k_- - k_- (N-N') k_z)
\cr
C^{\dag}  &=& -\frac{1}{3} (k_+ (N-N') k_z - k_z (N-N') k_+)
\end{array}
\end{equation}

\noindent with $L,M,N,N'$ being the Stravinou-van Dalen mass parameters. 

By setting the parameters constant, Eq.~(\ref{eq1}) Hamiltonian turns into the Luttinger Kohn Hamiltonian:
%
%\begin{equation}
%\label{eqb1}
% {\mathbb H}_{LK}=- \left( 
%\begin{array}{llll}
%\hat{P} + \hat{Q} & -\hat{S} & \hat{R} & 0 \\
%-\hat{S}^{\dag}& \hat{P} - \hat{Q} & 0 & \hat{R} \\ 
% \hat{R}^{\dag}& 0& \hat{P} - \hat{Q} & \hat{S}  \\
% 0& \hat{R}^{\dag}& \hat{S}^{\dag}& \hat{P} + \hat{Q}  
%\end{array}
%\right)
%\end{equation}
%\noindent with
%%
%\begin{equation}
%\label{eqb2}
%\begin{array}{lll}
%P &=& [\gamma_1 (k_x^2 + k_y^2 + k_z^2)]/2 \cr
%Q &=& [\gamma_2 (k_x^2 + k_y^2 - 2 k_z^2)]/2 \cr
%R &=& [-\sqrt{3} \gamma_2 (k_x^2 - k_y^2) + i \, 2 \sqrt{3} \gamma_3 k_x k_y]/ 2 \cr
%S &=& \gamma_3 \sqrt{3} (k_x - i \, k_y) k_z \cr
%S^{\dag} &=& \gamma_3 \sqrt{3} k_z (k_x + i  \, k_y) \cr
%R^{\dag}  &=& [-\sqrt{3} \gamma_2 (k_x^2 - k_y^2) - i \, 2 \sqrt{3} \gamma_3 k_x k_y]/2\cr
%Q^{\dag} &=& Q
%\end{array}
%\end{equation}
%
The Stravinou-van Dalen parameters are then related to the Luttinger parameters $\gamma_1,\gamma_2,\gamma_3$  as $L-M=-3 \gamma_2$, $3 L+M=-2 \gamma_1-5 \gamma_2$, $N-N'=1+ \gamma_1-2 \gamma_2-3 \gamma_3$ and $N+N'=-3 \gamma_3$.

${\mathbb H}_{B}^{[001]} $ represents the terms coming from a magnetic field applied along the growth ($z$) direction, $B$:\cite{PlanellesJPCM}
\begin{equation}
\label{eqr2}
{\mathbb  H}_B = - \left( \frac{B^2}{8} (x^2+y^2) +\frac{B}{2} (x k_y - y k_x) \right) \left( (\gamma_1-\frac{5}{2} \gamma_2) {\mathbb  I} +
                 \gamma_2 {\mathbb  J}_z^2 \right) + \kappa \mu_B B {\mathbb  J}_z.
\end{equation}
\noindent with ${\mathbb  I}$ a rank-4 identity matrix, ${\mathbb  J}_z$ the angular momentum z-component matrix ($J=3/2$),
$\kappa=4/3$ the effective g-factor\cite{bree} and $\mu_B$ the Bohr magneton. 

\begin{figure}[h]
\begin{center}
\includegraphics[width=0.45\textwidth]{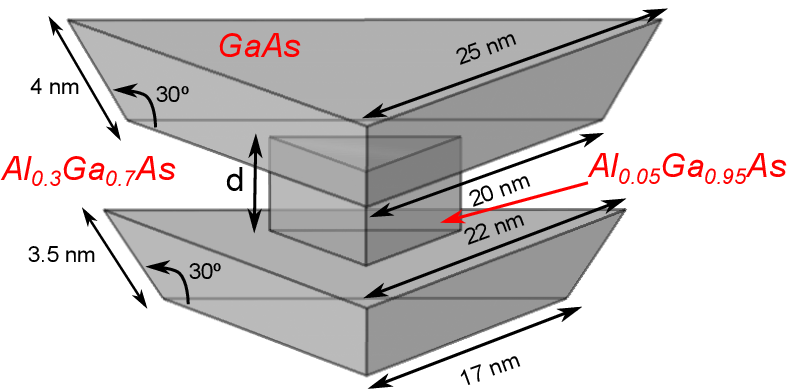}
\end{center}
\caption{QDM formed by triangular QDs. Vertical electric and magnetic fields are applied.}\label{fig1}
\end{figure}

${\mathbb H}_{strain}^{[001]}$ is the strain Hamiltonian, formally identical to Eq.~(\ref{eq1}) with the products $k_i\,k_j$ replaced by the strain tensor component $\epsilon_{ij}$.
$e$ and $\phi_{pz}$ are the hole charge and strain-induced piezoelectric potential -which is also diagonal-,  $F$ an axial electric field and  $V(\mathbf{r})$ the band-offset potential. For the triangular QDMs we use a structure like that obtained by metallorganic vapor deposition\cite{ZhuSMALL}. The structure is illustrated in Fig.~\ref{fig1}, using similar geometry and composition to the pyramidal QDMs of Refs.~\cite{MicheliniX,ZhuPRB}. Thus, the QDs are made of GaAs, the barrier of Al$_{0.3}$Ga$_{0.7}$As and the vertical wire connecting the dots of Al$_{0.05}$Ga$_{0.95}$As. We note that the vertical wire plays no critical role in the phenomena we describe. The robustness of the results are checked by carrying out additional
calculations with triangular round edges and also breaking the symmetry with two round and a
sharp edge (not shown).
For comparison, we also study QDMs made of hexagonal QDs. The sides of the hexagons are taken to have the same dimensions as those in Fig.~\ref{fig1}, and a hexagonal wire is used which preserves the $C_6$ symmetry.\\

For [111] grown QDs, the hole Hamiltonian becomes:
\begin{equation}
{\mathbb H}^{[111]} =  {\mathbb H}_{BF}^{[111]} + {\mathbb H}_{B}^{[111]} + {\mathbb H}_{strain}^{[111]} + \left( V(\mathbf{r})\, + \,e\,(\phi_{pz}^{[111]}(\mathbf{r}) - F\,z)\right) \,\mathbb{I}.
\label{eq:H111}
\end{equation}
\noindent where we have obtained ${\mathbb H}_{BF}^{[111]}$ and ${\mathbb H}_{strain}^{[111]}$ from Eq.~(\ref{equat1}) by writing $k_i$ and $\mathbb J_i$ appearing in the Hamiltonian as a function of  the new coordinates $k'_i$ and $\mathbb J'_i$ according to,
\begin{equation}
\label{rotmat2}
\mathbf k = \mathbb M \mathbf k' \;\;\;\;\;\;\;\;\;\; \mathbb J = \mathbb M \mathbb J'
\end{equation}
\noindent where $\mathbb M$ is the rotation matrix:\cite{XiaPRB}
\begin{equation}
\label{rotmat1}
\mathbb M=\left( 
\begin{array}{lll}
 \frac{1}{\sqrt{6}} & -\frac{1}{\sqrt{2}} & \frac{1}{\sqrt{3}} \\  
 \frac{1}{\sqrt{6}} & \frac{1}{\sqrt{2}} & \frac{1}{\sqrt{3}}  \\
 -\sqrt{\frac{2}{3}} & 0 & \frac{1}{\sqrt{3}} 
\end{array}
\right).
\end{equation}
\noindent Once we reach the new matrices, the prime is removed from $k_i$ for the sake of a better presentation. The Burt-Foreman Hamiltonian in the [111] direction then reads:
\begin{equation}
\label{eqd3}
 {\mathbb  H}_{BF}^{[111]}  =- \frac{1}{2} \left( 
\begin{array}{llll} 
\hat{P}' & -\hat{S_-} & \hat{R} & 0 \\ 
-\hat{S_-}^{\dag} & \hat{P"} & \hat{C} & \hat{R} \\ 
\hat{R}^{\dag} & \hat{C}^{\dag}&  \hat{P"}^* & \hat{S_+}^{\dag} \\ 
0 & \hat{R}^{\dag} & \hat{S_+} & \hat{P'}^*
\end{array}
\right)
\end{equation}
\noindent where,
\begin{equation}
\label{eqd4}
\begin{array}{lll}
\hat{P'} &=& \left[k_x (\gamma_1+\gamma_3) k_x + k_y (\gamma_1+\gamma_3) k_y + k_z (\gamma_1-2 \gamma_3) k_z\right]-  i \, \left[k_x (\gamma_1-2 \gamma_2-3 \gamma_3) k_y - k_y  (\gamma_1-2 \gamma_2-3 \gamma_3) k_x\right]
\cr
\hat{P"}  &=& \left[k_x (\gamma_1-\gamma_3) k_x + k_y (\gamma_1-\gamma_3) k_y + k_z (\gamma_1+2 \gamma_3) k_z\right] -\frac{i}{3} \, \left[k_x (\gamma_1-2 \gamma_2-3 \gamma_3) k_y - k_y  (\gamma_1-2 \gamma_2-3 \gamma_3) k_x\right]
\cr
\hat{S_-}  &=& -\frac{1}{\sqrt{3}} \left\{ (k_- \gamma_1 k_z- k_z \gamma_1 k_-)+\sqrt{2} \left[k_x (\gamma_2-\gamma_3) k_x - k_y (\gamma_2-\gamma_3) k_y + i \,
        (k_x (\gamma_2-\gamma_3) k_y+k_y (\gamma_2-\gamma_3) k_x)\right] \right. \cr 
        & & \left.      -2 \left[ 2k_- (\gamma_2+\gamma_3) k_z- k_z \gamma_3 k_-\right] \right\} 
\cr
\hat{S_+}  &=&  -\frac{1}{\sqrt{3}} \left\{ (k_+ \gamma_1 k_z- k_z \gamma_1 k_+)+\sqrt{2} \left[k_x (\gamma_2-\gamma_3) k_x - k_y (\gamma_2-\gamma_3) k_y - i \,
        (k_x (\gamma_2-\gamma_3) k_y+k_y (\gamma_2-\gamma_3) k_x)\right] \right. \cr 
        & & \left.      -2 \left[ 2k_+ (\gamma_2+\gamma_3) k_z- k_z \gamma_3 k_+ \right] \right\} 
\cr
\hat{R}  &=& - \frac{1}{\sqrt{3}}   \left\{k_- (\gamma_2+2\gamma_3) k_--\sqrt{2} \left[ k_z (\gamma_2-\gamma_3) k_++k_+ (\gamma_2-\gamma_3) k_z\right]\right\}
\cr
\hat{C}  &=&  -\frac{2}{3}  \left[ k_z (\gamma_1-2 \gamma_2-3 \gamma_3) k_- - k_- (\gamma_1-2 \gamma_2-3 \gamma_3) k_z\right]
\end{array}
\end{equation}

\noindent The ${\mathbb H}_{B}$ Hamiltonian rotation requires a special care. It was obtained in \cite{PlanellesJPCM} by initially disregarding the effect of the remote bands and later enclosing it, replacing the mass by the effective mass. Then, since we rotate the crystalline structure keeping the axes fixed and then the Bloch functions, the form of ${\mathbb H}_B$ does not change. However, the effect of the remote bands does change, so that one should replace $\gamma_2$ by $\gamma_3$  in the expression of the effectives masses. This modification must be introduced in ${\mathbb H}_{B}^{[001]}$ to reach ${\mathbb H}_{B}^{[111]}$.\\

To calculate the strain  $\epsilon_{ij}$ in ${\mathbb  H}_{strain}^{[111]}$, we take the elastic constants of $[001]$ grown heterostructures, $C^{[001]}$, and rotate the axes. The resulting elastic constants $C^{[111]}$ are related to $C^{[001]}$ by:
\begin{equation}
\label{eqr5}
C^{[111]}_{ijkl}=\sum_{a,b,c,d}\mathbb M_{ia}\mathbb M_{jb}\mathbb M_{kc}\mathbb M_{ld} C^{[001]}_{abcd}
\end{equation}
\noindent Likewise, for the piezoelectric potential we rotate the axes and obtain: 
\begin{equation}
\label{eqr6}
p_i=\sum_k e^{[111]}_{ijk}\epsilon_{jk}
\end{equation}
\noindent with $e^{[111]}_{ijk}=\sum_{a,b,c} \mathbb M_{ia}\mathbb M_{jb}\mathbb M_{kc} e^{[001]}_{abc}$.\\

Hamiltonians (\ref{eq:H}) and ($\ref{eq:H111}$) are solved numerically after obtaining the strain
tensors and piezoelectric fields using the \emph{Comsol} $4.2$ package. 
The material parameters of GaAs, AlAs and InAs are taken from Ref.~\cite{VurgaftmanJAP}, 
except for the crystal density, dielectric constant and piezoelectric coefficient, which are
obtained from Ref.~\cite{Madelung_book}.  
Linear interpolations are used for all alloys parameters. Luttinger parameters 
are inferred from the linearly interpolated masses. 

\section{Character table and product table for the double group $\bar {\mathbf{C}}_{3h}$}

\noindent The character table we use is:

\begin{equation}
\label{ch3_7a}
\begin{array}{l| l l l l l l l l l l l l|l}
\hline
\bar C_{3h}& E & C_3^+ & C_3^- & \sigma_h & S_3^+ & S_3^- &\bar E & \bar C_3^+ & \bar C_3^- & \bar \sigma_h & \bar S_3^+ & \bar S_3^-& {\rm basis}\cr
\hline
A^\prime &1& 1& 1& 1& 1& 1& 1& 1& 1& 1& 1& 1&   J_z   \cr
E_+^\prime &1& \omega& \omega^*& 1& \omega& \omega^*& 1& \omega& \omega^*& 1& \omega& \omega^*&   x + i y   \cr
E_-^\prime &1& \omega^*& \omega& 1& \omega^*& \omega& 1& \omega^*& \omega& 1& \omega^*& \omega&    x - i y   \cr
A^{\prime\prime} &1& 1& 1& -1& -1& -1& 1& 1& 1& -1& -1& -1&   z   \cr
E_+^{\prime\prime} &1& \omega& \omega^*& -1& -\omega& -\omega^*& 1& \omega& \omega^*& -1& -\omega& -\omega^*&    J_x + i J_y    \cr
E_-^{\prime\prime} &1& \omega^*& \omega& -1& -\omega^*& -\omega& 1& \omega^*& \omega& -1& -\omega^*& -\omega&     J_x - i J_y   \cr
\hline
E_{-1/2} &1& -\omega& \omega^*& i& -i \omega& i \omega^*& -1& \omega& -\omega^*& -i& i \omega& -i \omega^*&   J_{-1/2}   \cr
E_{1/2} &1& -\omega^*& \omega& -i& i \omega^*& -i \omega& -1& \omega^*& -\omega& i& -i \omega^*& i \omega&   J_{1/2}   \cr
E_{5/2} &1& -\omega& \omega^*& -i& i \omega& -i \omega^*& -1& \omega& -\omega^*& i& -i \omega& i \omega^*&  \cr % J_{5/2}   \cr
E_{-5/2} &1& -\omega^*& \omega& i& -i \omega^*& i \omega& -1& \omega^*& -\omega& -i& i \omega^*& -i \omega&    \cr %J_{-5/2}  \cr
E_{-3/2} &1& -1& 1& i& -i& i& -1& 1& -1& -i& i& -i&  \cr % J_{-3/2}    \cr
E_{3/2} &1& -1& 1& -i& i& -i& -1& 1& -1& i& -i& i&   \cr % J_{3/2}   \cr
\hline
\end{array}
\end{equation} \\

\noindent where $\omega=e^{i\,\frac{2\pi}{3}}$ and $J_i$ is the $i$-th component of the  angular momentum. \\ %$J=3/2$ angular momentum. 

\noindent The table of products of $\bar {\mathbf{C}}_{3h}$ irreducible  representations is:

\begin{equation}
\label{ch3_7}
\begin{array}{l| l l l l l l l l l l l l|}
  & A^\prime & E_+^\prime & E_-^\prime & A^{\prime\prime} & E_+^{\prime\prime} & E_-^{\prime\prime} & E_{-1/2} & E_{1/2} & E_{5/2} & E_{-5/2} & E_{-3/2} & E_{3/2} \cr
\hline
A^\prime  & A^\prime & E_+^\prime & E_-^\prime & A^{\prime\prime} & E_+^{\prime\prime} & E_-^{\prime\prime} & E_{-1/2} & E_{1/2} & E_{5/2} & E_{-5/2} & E_{-3/2} & E_{3/2} \cr
E_+^\prime  &     & E_-^\prime & A^\prime & E_+^{\prime\prime} & E_-^{\prime\prime} & A^{\prime\prime} & E_{-5/2} & E_{3/2} & E_{1/2} & E_{-3/2} & E_{-1/2} & E_{5/2} \cr
E_-^\prime  &     &     & E_+^\prime & E_-^{\prime\prime} & A^{\prime\prime} & E_+^{\prime\prime} & E_{-3/2} & E_{5/2} & E_{3/2} & E_{-1/2} & E_{-5/2} & E_{1/2} \cr
A^{\prime\prime}  &     &     &     & A^\prime & E_+^\prime & E_-^\prime & E_{5/2} & E_{-5/2} & E_{-1/2} & E_{1/2} & E_{3/2} & E_{-3/2} \cr
E_+^{\prime\prime}  &     &     &     &     & E_-^\prime & A^\prime & E_{1/2} & E_{-3/2} & E_{-5/2} & E_{3/2} & E_{5/2} & E_{-1/2} \cr
E_-^{\prime\prime}  &     &     &     &     &     & E_+^\prime & E_{3/2} & E_{-1/2} & E_{-3/2} & E_{5/2} & E_{1/2} & E_{-5/2} \cr
E_{-1/2}  &     &     &     &     &     &     & E_-^{\prime\prime} & A^\prime & E_-^\prime & A^{\prime\prime} & E_+^{\prime\prime} & E_+^\prime \cr
E_{1/2}  &     &     &     &     &     &     &     & E_+^{\prime\prime} & A^{\prime\prime} & E_+^\prime & E_-^\prime & E_-^{\prime\prime} \cr
E_{5/2}  &     &     &     &     &     &     &     &     & E_-^{\prime\prime} & A^\prime & E_+^\prime & E_+^{\prime\prime} \cr
E_{-5/2} &     &     &     &     &     &     &     &   &     & E_+^{\prime\prime} & E_-^{\prime\prime} & E_-^\prime \cr
E_{-3/2} &     &     &     &     &     &     &     &   &     &     & A^{\prime\prime} & A^\prime \cr
E_{3/2} &     &     &     &     &     &     &     &   &     &     &     & A^{\prime\prime} \cr
\hline
\end{array}
\end{equation} 

\section{Symmetry of the Hamiltonian and the wave functions}

The employed ${\mathbb H}_{BF}$ Hamiltonian is mass-position-dependent. Actually, in our system the mass parameters are constant within every domain, and have a sudden jump at the edge between neighboring domains. For the sake of easiness, to discuss on symmetry, we consider  the Luttinger-Kohn constant-mass parameters  ${\mathbb H}_{LK}$ limit of ${\mathbb H}_{BF}$.\\

\noindent The symmetry of the $ {\mathbb H}_{LK}$ Hamiltonian (eqs. \ref{eq1}, \ref{eqd3} employing constant mass parameters) including a triangular confining potential and an axial magnetic field is $C_3$. However, within the axial approximation\cite{Voon_book}  ($\gamma_2=\gamma_3$) it reaches $C_{3h}$. The symmetries of their matrix element operators can then be calculated  from the above character table and the expressions on eqs. \ref{eq2}, \ref{eqd4} (assuming constant mass parameters):
\begin{equation}
\label{eqc5}
\Gamma_{H_{LK}}=
\left( 
\begin{array}{llll}
A^{\prime} & E_-^{\prime \prime} & E_+^{\prime} & 0 \\ 
E_+^{\prime \prime} & A^{\prime} & 0 & E_+^{\prime } \\ 
E_-^{\prime} & 0& A^{\prime} & E_-^{\prime \prime} \\ 
0 & E_-^{\prime} & E_+^{\prime  \prime} & A^{\prime} 
\end{array} \right).
\end{equation}

\noindent Accordingly, the symmetry of the envelope bonding/anti-bonding ground state functions must be: 
\begin{equation}
\label{eqd5}
\left( 
\begin{array}{l}
A^{\prime}  \; (b)\\ 
E_+^{\prime \prime} \;(a)\\ 
E_-^{\prime } \;(b) \\ 
A^{\prime \prime} \;(a)
\end{array} \right) \; \; \; {\rm and}  \; \; \; 
\left( 
\begin{array}{l}
A^{\prime \prime} \; (a)\\ 
E_+^{\prime } \; (b)\\ 
E_-^{\prime \prime}  \; (a) \\ 
A^{\prime} \;(b)
\end{array} \right)
\end{equation}

\noindent where the labels $(a), (b)$ indicates the bonding/anti-bonding character of the components. These envelope components, combine with the Bloch functions yielding the wave function.\\

\noindent The Bloch functions are built as the symmetry-adapted product of $J=1$ angular momentum functions and the  $J=1/2$ spin  functions. However, the presence of the
mirror symmetry $\sigma_h$ allows to employ bonding ( $\chi (\sigma_h)=1$) and  anti-bonding ( $\chi (\sigma_h)=-1$) $J=1$ angular momentum functions.
For example, employing anti-bonding angular momentum functions we have:

\begin{equation}
\begin{array}{ll}
\label{app5}
|3/2,3/2\rangle =-\frac{1}{\sqrt{2}} |(X + i \, Y) \uparrow\rangle & |3/2,-3/2\rangle =\frac{1}{\sqrt{2}} |(X -i \, Y) \downarrow\rangle \cr
\cr
|3/2,1/2\rangle =\sqrt{\frac{2}{3}} |Z \uparrow\rangle - \frac{1}{\sqrt{6}}  |(X + i \, Y) \downarrow\rangle & 
                                    |3/2,-1/2\rangle =\sqrt{\frac{2}{3}} |Z \downarrow\rangle + \frac{1}{\sqrt{6}}  |(X - i \, Y) \uparrow\rangle  
\end{array}         
\end{equation}

\noindent The bonding  $J=1$ angular momentum functions are like the antibonding where $X, Y,$ and $Z$ are replaced by $J_x, J_y,$ and $J_z$. As a result, the table of products 
allows us to determine the symmetries of the Bloch functions:
\begin{equation}
\begin{array}{ll}
\label{app6}
|3/2,3/2^a\rangle \to E_{3/2} & |3/2,3/2^b\rangle \to E_{-3/2} \cr
|3/2,1/2^a\rangle \to E_{-5/2} & |3/2,1/2^b\rangle \to E_{1/2} \cr
|3/2,-1/2^a\rangle \to E_{5/2} & |3/2,-1/2^b\rangle \to E_{-1/2}\cr
|3/2,-3/2^a\rangle \to E_{-3/2}& |3/2,3/2^b\rangle \to E_{3/2}
\end{array}         
\end{equation}

 \noindent The Bloch functions symmetries required to combine with the envelope components eq. \ref{eqd5} (left) yielding $E_{3/2}$ and $E_{-3/2}$, and those required to combine with eq. \ref{eqd5} (right) yielding $E_{3/2}$ and $E_{-3/2}$, are:

\begin{equation}
\label{eqd25}
\left( 
\begin{array}{l}
E_{3/2}\\ 
E_{-5/2}\\ 
E_{5/2}\\ 
E_{-3/2}
\end{array} \right) \; \; \; {\rm and}  \; \; \; 
\left( 
\begin{array}{l}
E_{-3/2}\\ 
E_{1/2}\\ 
E_{-1/2}\\ 
E_{3/2}
\end{array} \right) \; \; ; \; \; \;
\left( 
\begin{array}{l}
E_{-3/2}\\ 
E_{1/2}\\ 
E_{-1/2}\\ 
E_{3/2}
\end{array} \right) \; \; \; {\rm and}  \; \; \; 
\left( 
\begin{array}{l}
E_{3/2}\\ 
E_{-5/2}\\ 
E_{5/2}\\ 
E_{-3/2}
\end{array} \right)
\end{equation}

\section{Spin mixing in QDMs grown along [001]}

\begin{figure}[h]
\begin{center}
\includegraphics[width=0.45\textwidth]{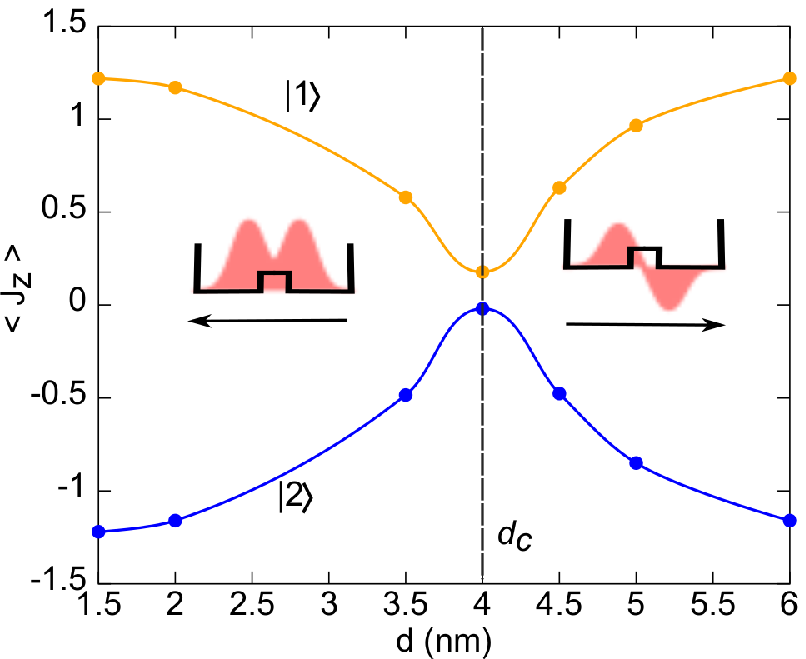}
\caption{Calculated spin purity of states $|1\rangle$ and $|2\rangle$ for different
interdot distances at the resonant electric field for a triangular QDM grown along [001].}\label{fig2}
\end{center}
\end{figure}

In the paper we have considered triangular QDs grown along the [111] direction
because they are formed naturally in that direction. The physics leading to hole spin mixing
is however connected with the envelope symmetry, and does not depend on the crystal orientation.
To illustrate this point, in Fig.~\ref{fig2} we plot the expectation value of the Bloch angular
momentum $\langle J_z \rangle$ as a function of the interdot distance $d$ for the upper 
Zeeman doublet of GaAs/Al$_{0.3}$Ga$_{0.7}$As  QDMs (states $|1\rangle$ and $|2\rangle$). 
All the parameters are taken as in Fig.~3(d) of the paper, except that now the QDM is grown 
along [001] -i.e. Hamiltonian $\mathbb{H}^{[001]}$ instead of $\mathbb{H}^{[111]}$-.
As can be seen, the picture is qualitatively the same as that obtained for [111] grown QDMs
(Fig.3(d) in the paper). The main difference is that the critical distance $d_c$, where the
bonding-antibonding reversal takes place and hole spin mixing is maximized, is now 
shifted towards longer interdot distances. This is because the effective masses of HH
along [001] are lighter than those along [111].
Therefore, the tunneling is stronger. 

\noindent Indeed, if we take constant mass parameters in 
${\mathbb H}_{BF}^{[001]}$, eq. \ref{eq1}, we obtain
the corresponding Luttinger-Kohn Hamiltonian:

\begin{equation}
\label{eqb1}
 {\mathbb H}_{LK}^{[001]}=- \left( 
\begin{array}{llll}
\hat{P} + \hat{Q} & -\hat{S} & \hat{R} & 0 \\
-\hat{S}^{\dag}& \hat{P} - \hat{Q} & 0 & \hat{R} \\
\hat{R}^{\dag}& 0& \hat{P} - \hat{Q} & \hat{S}  \\
0& \hat{R}^{\dag}& \hat{S}^{\dag}& \hat{P} + \hat{Q} 
\end{array} 
\right)
\end{equation}
\noindent with
\begin{equation}
\label{eqb2}
\begin{array}{lll}
\hat{P} \pm \hat{Q} &=& [(\gamma_1 \pm \gamma_2) (k_x^2 + k_y^2) + (\gamma_1 \mp 2\gamma_2) k_z^2)]/2 \cr
\hat{R} &=& [-\sqrt{3} \gamma_2 (k_x^2 - k_y^2) + i \, 2 \sqrt{3} \gamma_3 k_x k_y]/ 2 \cr
\hat{S} &=& \gamma_3 \sqrt{3} (k_x - i \, k_y) k_z \cr
\end{array}
\end{equation}

\noindent Also, if we take constant mass parameters in ${\mathbb H}_{BF}^{[111]}$, eq. \ref{eqd3}, we obtain:

\begin{equation}
\label{eqb1b}
 {\mathbb H}_{LK}^{[111]}=-\frac{1}{2} \left( 
\begin{array}{llll}
\hat{P} + \hat{Q} & -\hat{S} & \hat{R} & 0 \\
-\hat{S}^{\dag}& \hat{P} - \hat{Q} & 0 & \hat{R} \\
\hat{R}^{\dag}& 0& \hat{P} - \hat{Q} & \hat{S}  \\
0& \hat{R}^{\dag}& \hat{S}^{\dag}& \hat{P} + \hat{Q} 
\end{array} 
\right)
\end{equation}
\noindent with
\begin{equation}
\label{eqb2b}
\begin{array}{lll}
\hat{P} \pm \hat{Q} &=& (\gamma_1 \pm \gamma_3) (k_x^2 + k_y^2) + (\gamma_1 \mp 2 \gamma_3) k_z^2 \cr
\hat{R} &=& -\frac{1}{\sqrt{3}} (\gamma_2+2 \gamma_3) k_-^2  + \frac{2\sqrt{2}}{\sqrt{3}} (\gamma_2-\gamma_3) k_+ k_z \cr
\hat{S} &=& -\frac{\sqrt{2}}{\sqrt{3}} (\gamma_2-\gamma_3)  k_+^2 + \frac{2}{\sqrt{3}} \,(2 \gamma_2+\gamma_3) k_- k_z \cr
\end{array}
\end{equation}

\noindent By comparing $\hat{P}+\hat{Q}$ in Eq.~(\ref{eqb2}) with Eq.~(\ref{eqb2b}), one can note the
different HH effective masses in the z direction: $1/(\gamma_1-2\gamma_2)$ vs $1/(\gamma_1-2\gamma_3)$.

\noindent It is worth noting that the $\hat{R}$ operator in ${\mathbb H}_{LK}^{[001]}$ does not have $C_3$ rotational 
symmetry (the $C_3$ character table can be obtained from that of $C_{3h}$ by just considering rotations and 
keeping rows 1-3,7-8 and 12).
However, for many III-V materials including GaAs,  $\gamma_2 \approx \gamma_3$ and one can approximate 
them both by $ \bar \gamma=(\gamma_2+\gamma_3)/2$ in the $\hat{R}$ and  $\hat{R}^{\dag}$ matrix elements, 
thus yielding $\mathbb{H}_{LK}^{[001]}$ with axial symmetry,\cite{Voon_book} that is reduced to $C_3$ 
(or  $C_{3h}$) symmetry by the triangular confining potential and the magnetic field. 
On the other hand, Hamiltonian ${\mathbb H}_{LK}^{[111]}$ does have $C_3$ symmetry, but --as discussed
in the paper-- the axial approximation in both $R$ and $S$ matrix elements is needed to display exact 
$C_{3h}$ symmetry.

In short, in both crystallographic directions the Hamiltonian has approximate $C_{3h}$ symmetry,
which becomes exact if the axial approximation is assumed.
Similar considerations on the axial approximation hold for the strain terms. 

\section{Spin mixing in InAs/GaAs QDMs}

Next, we consider InAs/GaAs QDMs grown along [001], similar to those obtained by self-assembled growth\cite{StinaffSCI}
but with triangular (pyramidal) QD shape. Unlike for GaAs/AlGaAs, no interdot wire is present in this case. 
On the other hand, strain and piezoelectricity now play a significant role. 
 
\begin{figure}[h]
\begin{center}
\includegraphics[width=0.45\textwidth]{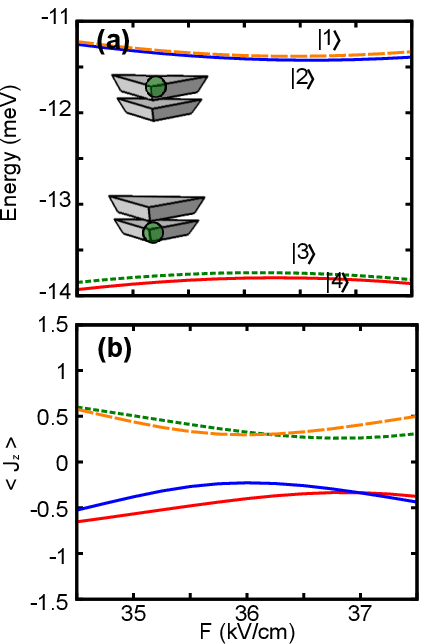}
\caption{(a) and (b): hole energy levels and Bloch angular momentum expectation value of a triangular InAs/GaAs QDM
grown along [001], as a function of a vertical electric field.  
The insets in (a) show the hole localization for each doublet.}\label{fig3}
\end{center}
\end{figure}

Figure \ref{fig3} shows the spin purity of the four fist hole states as a function of an external electric
field. One can see that also in this case there is a strong spin mixing ($|\langle J_z \rangle| \ll 3/2$).
An inspection of the individual spinor components (not shown) reveals that also for InAs/GaAs LH components
play a minor role. Most of the mixing in Fig.~\ref{fig3} comes from admixture between HH $J_z=+3/2$ and $J_z=-3/2$ 
components, following the symmetry-induced mechanism described in our paper.

It is remarkable that the spin mixing here takes place for bonding and antibonding states split by more than 2 meV.
Besides, the spin mixing takes place over a wide window of electric fields (wider than for GaAs).
This is because of the stronger SOI of InAs as compared to GaAs.\cite{gamma2} These results suggest that the eventual 
design of triangular InAs/GaAs QDMs would also form a promising
system for hole spin manipulation.

\section{Perturbational estimate of the spin mixing strength}

In this section we expand the discussion of Fig.~3 based on perturbation theory. 
The goal is to show that the spin in triangular QDMs is much stronger than in circular QDMs.

The Hamiltonian describing the hole states in a QDM, be it $\mathbb H^{[001]}$ or $\mathbb H^{[111]}$,
can be split as:
\begin{equation}
\mathbb{H}=\mathbb{H}_0 + \mathbb{H}^\prime. 
\end{equation}
\noindent Here $\mathbb{H}_0$ are the diagonal terms, whose eigenfunctions
are single-band HH or LH states:
\begin{eqnarray}
\label{states1}
|\Psi_{HH \Uparrow}^0(X^p)\rangle = f_{3/2}(X^p) \left(\begin{matrix} 1\cr 0 \cr 0\cr 0\end{matrix}\right) \; &,& \; |\Psi_{LH \Uparrow}^0(X^p)\rangle = f_{1/2}(X^p) \left(\begin{matrix} 0\cr 1 \cr 0\cr 0\end{matrix}\right) \cr
\cr
 |\Psi_{LH \Downarrow}^0(X^p)\rangle = f_{-1/2}(X^p) \left(\begin{matrix} 0\cr 0 \cr 1\cr 0\end{matrix}\right) \;  &,& \; |\Psi_{HH \Downarrow}^0(X^p)\rangle = f_{-3/2}(X^p) \left(\begin{matrix} 0\cr 0 \cr 0\cr 1\end{matrix}\right)
\end{eqnarray}
\noindent where $f_{J_z}(X^p)$ is the envelope function of $X$ rotational symmetry and $p$ parity. 
In turn, $\mathbb{H}^\prime$ represents the off-diagonal terms of $\mathbb H$, 
coming from $\mathbb{H}_{BF}$ and $\mathbb{H}_{strain}$. This term is responsible 
for the band coupling. Without loss of generality, because the symmetry of 
$\mathbb{H}_{BF}$ and $\mathbb{H}_{strain}$ is the same, in what follows we
consider GaAs/AlGaAs QDMs, where $\mathbb{H}_{strain}$ is negligible.
The analysis is further simplified replacing $\mathbb{H}_{BF}$ by its constant
mass analogue, $\mathbb{H}_{LK}$, Eqs.~(\ref{eqb1}) or (\ref{eqb1b}) within axial approximation. This
leads to the following expression for $\mathbb{H}^\prime$:

\begin{equation}
\mathbb{H}^\prime=-\left(\begin{matrix} 0 & -S & R &0 \cr -S^{\dagger} & 0 & 0 & R \cr R^{\dagger} & 0 & 0 & S \cr 0& R^{\dagger}& S^{\dagger} &0\end{matrix}      \right)
\end{equation}

\noindent Within this approximation the matrix element operators $R$ and $S$ are just proportional to
$k_-^2$ and $k_- k_z$, respectively.

\noindent Considering $\mathbb H'$ as a perturbative term, the mixing of states up to second order is given by:
\begin{equation}
\label{pert1}
|\Psi_k^{(2)}\rangle =\sum_{i\neq k} \left(\sum_{j\neq k} \frac{\langle \Psi_i^{(0)}|\mathbb{H}^\prime|\Psi_j^{(0)}\rangle}{E_k^0-E_i^0}
\frac{\langle \Psi_j^{(0)}|\mathbb{H}^\prime|\Psi_k^{(0)}\rangle}{E_k^0-E_j^0} \right) |\Psi_i^{(0)}\rangle 
\end{equation}

\noindent Note that the coupling between $|\Psi_{HH \Uparrow}^0(X_i^{p_i})\rangle$ and $|\Psi_{HH \Downarrow}^0(X_k^{p_k})\rangle$ via $\mathbb{H}^\prime$ 
requires $|\Psi_{LH \Uparrow}^0(X_j^{p_j})\rangle$ as intermediate state, yielding the contribution:
\begin{equation}
\label{pert2}
-\frac{\langle X_i^{p_i}|S|X_j^{p_j}\rangle \langle X_j^{p_j}|R| X_k^{p_k}\rangle}{\Delta E_{hh} \Delta E_{lh}}
\end{equation}
\noindent  where $\Delta E_{hh}$ is the energy difference
between the HH states  $|\Psi_{HH \Downarrow}^0(X_k^{p_k})\rangle$ and $|\Psi_{HH \Uparrow}^0(X_i^{p_i})\rangle$,
and $\Delta E_{lh}$ that between $|\Psi_{HH \Downarrow}^0(X_k^{p_k})\rangle$
and the  LH state $|\Psi_{LH \Uparrow}^0(X_j^{p_j})\rangle$.

\noindent Alternatively, this coupling can also be achieved with $|\Psi_{LH \Downarrow}^0(X_j^{p_j})\rangle$ as intermediate state,
yielding the contribution:

\begin{equation}
\label{pert3}
\frac{\langle X_i^{p_i}|R|X_j^{p_j}\rangle \langle X_j^{p_j}|S| X_k^{p_k}\rangle}{\Delta E_{hh} \Delta E_{lh}}.
\end{equation}

\noindent The matrix elements in the numerator of Eqs.~(\ref{pert2}), (\ref{pert3}) determine 
the selection rules in the band coupling process. E.g. in the $C_{3h}$ group, 
the matrix element operator $S$ has $E_-^{\prime\prime}$
symmetry ($R$ has  $E_+^{\prime}$), see Eq.~(\ref{eqc5}).
Then, a totally symmetric  $A^{\prime}$  $\langle\Psi_{HH \Uparrow}^0|$ ground state
must couple, via $S$, with a $|\Psi_{LH \Uparrow}^0\rangle$ state of symmetry $E_+^{\prime\prime}$
 ($A^{\prime} \otimes E_-^{\prime\prime} \otimes E_+^{\prime\prime} = A^{\prime}$,
otherwise the integral is zero). Next,  $\langle \Psi_{LH \Uparrow}^0|$ with
symmetry $E_-^{\prime\prime}$ (the complex conjugate of that of $|\Psi_{LH \Uparrow}^0\rangle$) 
must couple, via $R$, with  $|\Psi_{HH \Downarrow}^0\rangle$ of $A^{\prime\prime}$
symmetry ($ E_-^{\prime\prime} \otimes E_+^{\prime} \otimes A^{\prime\prime} = A^{\prime}$,
otherwise the integral is zero).
Likewise,  $\langle\Psi_{HH \Uparrow}^0|$ of  $A^{\prime}$ symmetry (note that $A^\prime$
and $A^{\prime \prime}$ are reals and then coincide with their complex conjugates) can 
couple via $R$ with  $|\Psi_{LH \Downarrow}^0\rangle$ of $E_-^{\prime}$ symmetry. Then,
$\langle \Psi_{LH \Downarrow}^0|$ of $E_+^{\prime}$ symmetry will couple, via $S$,
with $|\Psi_{HH \Downarrow}^0\rangle$ of $A^{\prime\prime}$ symmetry
($ E_+^{\prime} \otimes E_-^{\prime\prime} \otimes A^{\prime \prime} = A^{\prime}$).

\begin{figure}[h]
\begin{center}
\includegraphics[width=0.5\textwidth]{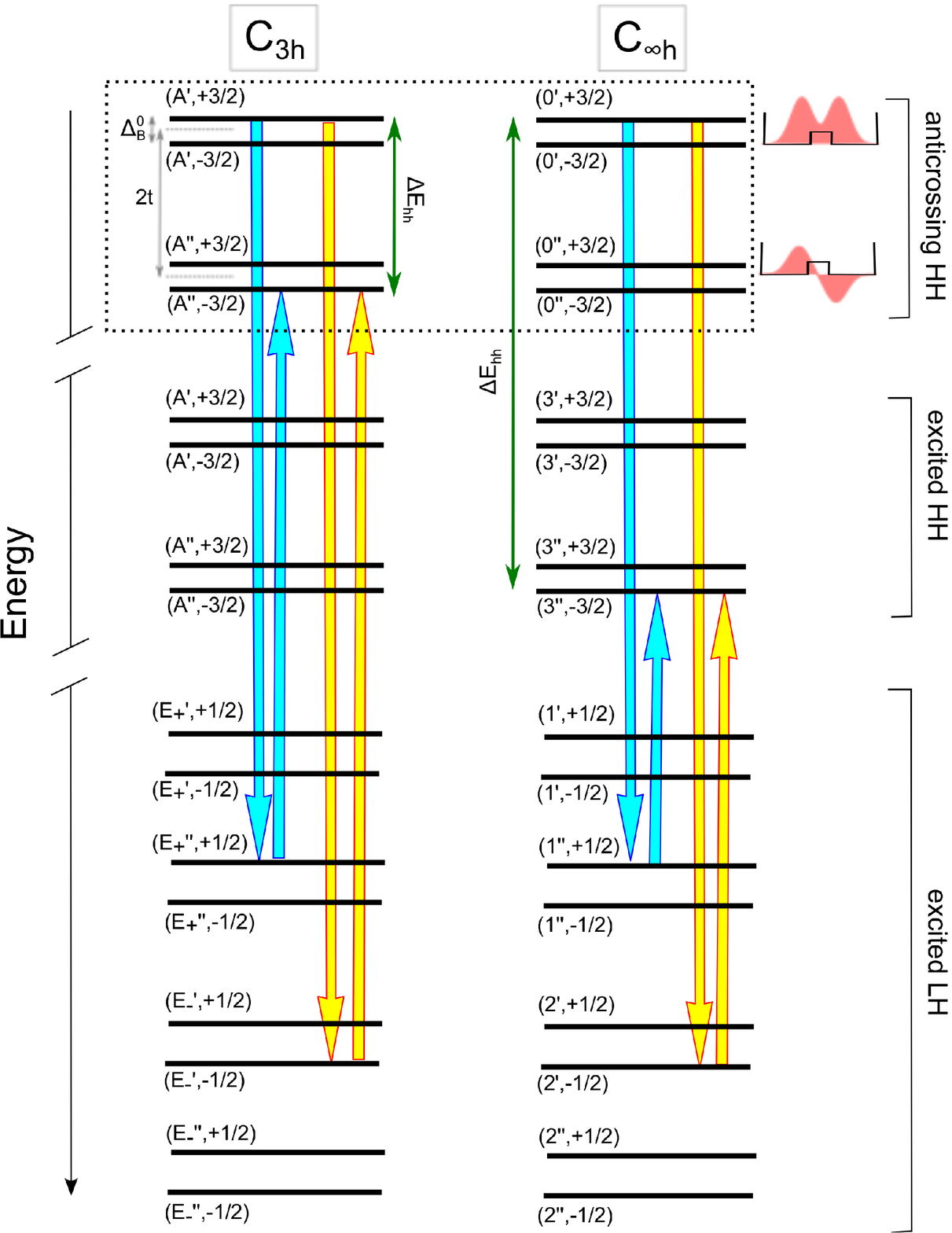}
\caption{Diagram of single band hole energy levels under a longitudinal magnetic field for
a QDM with symmetry $C_{3h}$ (left column) and $C_{\infty h}$ (right column). The labels
$(X^p,J_z)$ indicate the symmetry of the levels. $X$ represents the rotational
symmetry of the envelope function ($A$, $E_+$ and $E_-$ in $C_3$ and $M_z$ in $C_{\infty h}$),
$p=\prime$ or $\prime \prime$ represents the even/odd parity and $J_z$ indicates the non-zero component
of the spinor (e.g. $J_z=3/2$ means that the non-zero component is the first one,  $J_z=1/2$ the second, etc.) 
Thick arrows denote the symmetry allowed level couplings for
the ground state, as inferred from  Eqs.~(\ref{pert2}), (\ref{pert3}).}
\label{fig4}
\end{center}
\end{figure}

The above reasonings lead us to define the $C_{3h}$ allowed couplings, represented by thick 
vertical lines on the left side of Fig.~\ref{fig4}. Blue and yellow arrows correspond to either
contribution, eqs. (\ref{pert2}) and (\ref{pert3}). In the figure we have simplified the notation 
of the states in eq.~(\ref{states1}) to $(X^p,J_z)$, where $X$ is the rotational symmetry of the envelope function 
($A$, $E_+$ and $E_-$ in $C_3$ or $M_z$ in $C_{\infty}$), $p= \prime$ or $\prime \prime$ represents
the even/odd parity, and $J_z$ indicates the non-zero component of the four-fold spinor. % (e.g.
%$J_z=3/2$ mean that only the first component is non-zero, $J_z=-1/2$ taht the non-zero is the third one, etc.)

\noindent One can now compare with the case of circular QDMs, where the group is $C_{\infty h}$ 
and the envelope functions are labeled by $M_z$ and parity. 
In this group, the matrix element operator $R$ ($S$) is even (odd) and has $M_z=-2$ ($M_z=-1$).
Now  the $\langle\Psi_{HH \Uparrow}^0|$ ground state of $0^{\prime}$ symmetry can couple via $S$ ($-1^{\prime\prime}$)
with a $|\Psi_{LH \Uparrow}^0\rangle$ of symmetry $1^{\prime\prime}$. Then,  $\langle \Psi_{LH \Uparrow}^0|$
of symmetry  $-1^{\prime\prime}$ couple, via $R$ ($-2^{\prime}$), with  $|\Psi_{HH \Downarrow}^0\rangle$ of
$3^{\prime\prime}$ symmetry (because $-1-2+3=0$ and $\prime\prime\otimes\prime\otimes \prime\prime=\prime$).

\noindent $\langle\Psi_{HH \Uparrow}^0|$ $0^{\prime}$ may also couple, via $R$ ($-2^{\prime}$) with  
$|\Psi_{LH \Downarrow}^0\rangle$ $2^{\prime}$. In turn,  $\langle\Psi_{LH \Downarrow}^0|$ of $-2^{\prime}$
symmetry couples, via $S$ ($-1^{\prime\prime}$), with $|\Psi_{HH \Downarrow}^0\rangle$ of 
$3^{\prime\prime}$ symmetry.

\noindent Taking into account Fig.~\ref{fig4} and  Eqs.~(\ref{pert2}) and (\ref{pert3}) we can see that 
$\Delta E_{hh}$ in the denominator involved in $C_{3h}$ is quite smaller than that involved in 
 $C_{\infty h}$  and, therefore, the interaction should be much stronger.

\noindent As a matter of fact, the strong spin mixing at the resonant electric fields is a singular behavior of triangular QDMs.
In $C_{nh}$ symmetries with $n>3$ the spinor fourth component ($J_z=-3/2$) has different rotational symmetry symmetry than 
the first one ($J_z=3/2$), as in the  above discussed case of $C_{\infty h}$. 
Then, for similar reasons, the coupling between the states belonging to the first two doublets are also forbidden.
In the case of $C_{2h}$ QDM the symmetries of $S$ and $R$ are $B_g$ and $A_g$ respectively. Bonding (anti-bonding)
molecular orbitals are of $Ag$ ($B_u$) symmetry. Then, the doublet antibonding ground state (bonding first excited state) are 
$B_u \Uparrow$, $B_u \Downarrow$ ($A_g \Uparrow$, $A_g \Downarrow$). Therefore, any coupling among these four states
is forbidden by symmetry, as can be easily checked with the help of Eqs.~(\ref{pert2}) and (\ref{pert3}). 

%\bibliography{holetrianglebib}{}
%\bibliographystyle{unsrt}